\documentclass
[aps,prd,twocolumn,showpacs,superscriptaddress]
{revtex4-1}
\usepackage[svgnames]{xcolor}
\usepackage[colorlinks, linkcolor=Navy, urlcolor=Navy, citecolor=Navy]{hyperref}
\usepackage[all]{hypcap}
\newcommand{\pipi}{\pi^0\pi^0}
\newcommand{\psip}{\psi(3686)}

\newcommand{\MeVcc}{MeV/\textit{c}$^2$}
\newcommand{\GeVcc}{GeV/\textit{c}$^2$}

\newcommand{\BESIIIorcid}[1]{\href{https://orcid.org/#1}{\hspace*{0.1em}\raisebox{-0.45ex}
{\includegraphics[width=1em]{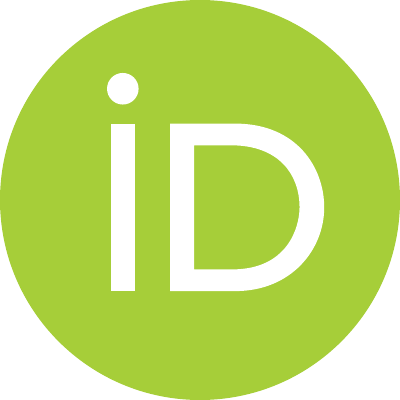}}}}

\usepackage{amsmath}
\usepackage[pdftex]{graphicx}
\usepackage{longtable}
\usepackage{multirow}
\usepackage{color}
\usepackage{braket}

\usepackage{lipsum}
\begin{document}

\title{Amplitude analysis of {\boldmath $\psip$} radiative decays into {\boldmath $\pipi$} system}

\author{
\begin{small}
 \begin{center}
M.~Ablikim$^{1}$\BESIIIorcid{0000-0002-3935-619X},
M.~N.~Achasov$^{4,c}$\BESIIIorcid{0000-0002-9400-8622},
P.~Adlarson$^{83}$\BESIIIorcid{0000-0001-6280-3851},
X.~C.~Ai$^{89}$\BESIIIorcid{0000-0003-3856-2415},
C.~S.~Akondi$^{31A,31B}$\BESIIIorcid{0000-0001-6303-5217},
R.~Aliberti$^{39}$\BESIIIorcid{0000-0003-3500-4012},
A.~Amoroso$^{82A,82C}$\BESIIIorcid{0000-0002-3095-8610},
Q.~An$^{78,65,\dagger}$,
Y.~H.~An$^{89}$\BESIIIorcid{0009-0008-3419-0849},
M.~S.~Anderson$^{39}$\BESIIIorcid{0009-0008-1550-2632}
% ...
\\
\vspace{0.2cm}
(BESIII Collaboration)\\
\vspace{0.2cm} {\it
$^{1}$ Institute of High Energy Physics, Beijing 100049, People's Republic of China\\
% ...
$^{4}$ Budker Institute of Nuclear Physics SB RAS (BINP), Novosibirsk 630090, Russia\\
% ...
$^{31}$ INFN Sezione di Ferrara, (A)INFN Sezione di Ferrara, I-44122, Ferrara, Italy; (B)University of Ferrara, I-44122, Ferrara, Italy\\
% ...
$^{39}$ Johannes Gutenberg University of Mainz, Johann-Joachim-Becher-Weg 45, D-55099 Mainz, Germany\\
% ...
$^{65}$ State Key Laboratory of Particle Detection and Electronics, Beijing 100049, Hefei 230026, People's Republic of China\\
% ...
$^{78}$ University of Science and Technology of China, Hefei 230026, People's Republic of China\\
% ...
$^{82}$ University of Turin and INFN, (A)University of Turin, I-10125, Turin, Italy; (B)University of Eastern Piedmont, I-15121, Alessandria, Italy; (C)INFN, I-10125, Turin, Italy\\
$^{83}$ Uppsala University, Box 516, SE-75120 Uppsala, Sweden\\
% ...
$^{89}$ Zhengzhou University, Zhengzhou 450001, People's Republic of China\\

\vspace{0.2cm}
$^{\dagger}$ Deceased\\
% ...
$^{c}$ Also at the Novosibirsk State University, Novosibirsk, 630090, Russia\\
% ...

}

 \end{center}
\end{small}
}

\date{October 12, 2026}

\begin{abstract}
       A model-independent amplitude analysis of the $\pipi$ system produced in
       radiative $\psip$ decays is presented. The analysis is based on the unique dataset
       of $(2712.4\pm14.3)\times10^6$ $\psip$ decays collected by the BESIII detector.
       Scalar and tensor partial-wave contributions are obtained
       as piecewise functions of $M_{\text{inv}}(\pipi)$ with minimal model assumptions.
       The observed contribution of the scalar partial wave exhibits similar behavior
       to the one reported in the study of radiative $J/\psi$ decays,
       while the tensor contributions provide no evidence for resonance-like
       structures except for the clearly visible $f_2(1270)$.
       The results are presented in the form of log-likelihood scans
       in $M_{\text{inv}}(\pipi)$ bins, which preserve all the statistical information and
       account for relevant systematic uncertainties.
       The model-independent description provided in this paper
       allows to combine the presented results with other related data in the
       development of phenomenological models.
\end{abstract}

\maketitle

\section{Introduction \label{sec:intro}}
       The search for the particles beyond the conventional quark model is
       one of the most important tasks in modern Quantum
       Chromodynamics (QCD)~\cite{physics_of_glueballs}.
       Among the most prominent candidates are tetraquarks, pentaquarks, hybrids and glueballs.
       While many tetraquark and pentaquark candidates have been reported
       since the development of QCD,
       glueballs -- color-singlet bound states composed purely of gluons -- still remain elusive for the most part.

       A wide range of theoretical predictions for the glueball mass spectra has been obtained so far.
       In particular, glueball masses were calculated within QCD Hamiltonian 
       approaches~\cite{glueball_masses_qpgb},
       QCD sum rules~\cite{glueball_masses_QCD_sum_rules},
       Dyson-Schwinger equations~\cite{glueball_masses_Dyson_Schwinger_1,
       glueball_masses_Dyson_Schwinger_2} and a graviton-soft-wall model under
       the AdS/QCD approach~\cite{glueball_masses_AdS_QCD}. Lattice calculations
       were also used extensively, both in quenched~\cite{glueball_mass_spectrum_quenched_lattice_1,
       glueball_mass_spectrum_quenched_lattice_2} and
       unquenched~\cite{glueball_mass_spectrum_unquenched_lattice} formulations.
       Although the predicted masses differ quantitatively between models,
       all of the predictions conclude that scalar, tensor and
       pseudoscalar glueballs are the three lightest glueball states, all with masses below $3$~\GeVcc.
       Lattice QCD calculations additionally predict glueball production rates in radiative $J/\psi$
       decays~\cite{lattice_scalar_glueball_yield, lattice_tensor_glueball_yield,
       lattice_pseudoscalar_glueball_yield}, suggesting that in radiative $J/\psi$ decays
       the tensor glueball yield is enhanced by a factor three relative to the scalar one.

       Glueball states have also been extensively searched for in various experiments
       (for example, see Review~\cite{50yearsQCD}). Among recent studies,
       the coupled channel analysis~\cite{scalar_glueball} of the BESIII data of radiative
       $J/\psi$ decays into $\pi \pi$, $K\bar{K}$, $\eta\eta$ and $\omega \phi$ final states, 
       combined with additional data, reported an enhancement interpreted as a scalar glueball candidate with  
       $M_G \approx 1865$~\MeVcc.
       However, an alternative description of the same spectra was obtained in a similar analysis 
       performed without constraints on the two-meson rescattering amplitudes~\cite{Rodas:2022aa}.
       The partial-wave analysis~\cite{X2370} of the BESIII data of radiative
       $J/\psi \to \gamma K_S^0 K_S^0 \eta^\prime$ decays measured the mass and width of the $X(2370)$
       state and its spin parity, all of which are consistent with the predictions for the lightest
       pseudoscalar glueball. In the study~\cite{Tensor_glueball_search},
       an enhancement in the tensor wave is observed, but with lower mass and a substantially
       lower yield than predicted for the tensor glueball. One possible interpretation proposed by 
       the authors is that this enhancement corresponds to the low-mass tail of a higher-mass 
       tensor glueball state. Such a scenario can be further investigated in radiative $\psip$ decays.

       The BESIII experiment and its data are particularly well-suited for glueball searches.
       The $c\bar{c}$-annihilation in charmonium decays allows for the creation of a glue-rich environment,
       which greatly enhances glueball production rates~\cite{physics_of_glueballs}.
       At the same time, the BESIII experiment has gathered the largest unique datasets
       of $J/\psi$ and $\psip$ decays in the world. Radiative decays into two-meson final states 
       are especially suitable for model-independent amplitude analyses.

       Radiative $J/\psi$ decays into the $\pipi$ system were previously studied
       by the BESIII collaboration~\cite{PWA_Jpsi_pipi}. However, the limited phase space in 
       $J/\psi$ decays restricts sensitivity to the high invariant-mass region, particularly 
       between $2.5$ and $3$~\GeVcc, which is of particular interest for tensor glueball searches.
       In contrast, radiative $\psip$ decays provide access to $\pipi$  invariant masses 
       from threshold up to the $\psip$ mass, enabling a more extensive study of the light-hadron spectrum. 
       Motivated by existing glueball mass predictions, the present analysis focuses primarily 
       on the mass region below $3$~\GeVcc. It is worth noting that the study of $\psip \to \gamma \pipi$ 
       decays is experimentally advantageous compared to the study of $\psip \to \gamma \pi^+ \pi^-$
       decays because of the latter channel's difficult to suppress background 
       from $\psip \to \pi^0 \pi^{+} \pi^{-}$ decays, with one additional photon in the final state. 
       Such backgrounds are absent in the neutral channel which is studied here. 
       Furthermore, $\psip \to \pi^0 \pi^{0} \pi^{0}$ decays are forbidden due to the conservation of $C$-parity.

\section{The BESIII detector \label{sec:bes3}}

       The BESIII detector~\cite{BESIII_Design} records symmetric $e^+e^-$ collisions
       provided by the BEPCII storage ring~\cite{Yu:IPAC2016-TUYA01}
       in the center-of-mass energy range from 1.84 to 4.95~GeV~\cite{ref:BEPCII_max_en},
       with a peak luminosity of $1.1 \times 10^{33}\;\text{cm}^{-2}\text{s}^{-1}$
       achieved at $\sqrt{s} = 3.773\;\text{GeV}$.
       BESIII has collected large data samples in this energy 
       region~\cite{BESIII_Physics_Programme, EcmsMea, EventFilter}.
       The cylindrical core of the BESIII detector covers 93\% of the full solid angle and consists of
       a helium-based multilayer drift chamber~(MDC), a time-of-flight system~(TOF), and
       a CsI(Tl) electromagnetic calorimeter~(EMC), which are all enclosed in a superconducting
       solenoidal magnet providing a 1.0~T magnetic field.
       The solenoid is supported by an
       octagonal flux-return yoke with resistive plate counter muon
       identification modules interleaved with steel.
       The charged-particle momentum resolution at $1~{\rm GeV}/c$ is $0.5\%$, and the
       ${\rm d}E/{\rm d}x$ resolution is $6\%$ for electrons from Bhabha scattering.
       The EMC measures photon energies with a resolution of $2.5\%$ ($5\%$) at $1$~GeV
       in the barrel (end cap) region. The time resolution in the plastic scintillator TOF barrel region
       is 68~ps, while that in the end cap region was 110~ps.
       The end cap TOF system was upgraded in 2015 using
       multigap resistive plate chamber technology, providing a time resolution of 60~ps,
       which benefits 83\% of the data used in this analysis~\cite{Li2017, Guo2017, Liu2022}.

\section{Data sets \label{sec:data_sets}}

       This analysis uses a dataset corresponding to $2712$ million $\psip$ events collected by 
       the BESIII detector during 2009, 2012 and 2021 at a center-of-mass energy ($E_\text{CMS}$) 
       of 3686~MeV. In addition, the dataset of $e^+e^-$ collisions with $E_\text{CMS} = 3650$~MeV is used.
       Detailed information on the datasets is provided in Ref.~\cite{Ablikim_2024}.

       Monte Carlo (MC) simulated data samples produced
       with a {\sc geant4}-based~\cite{geant4} software package, which
       includes the geometric description of the BESIII detector and
       the detector response, are used to determine detection efficiencies
       and to estimate backgrounds. The simulation models the beam energy spread
       and initial state radiation (ISR) in the $e^+e^-$ annihilations
       with the generator {\sc kkmc}~\cite{ref:kkmc}.
       The inclusive MC sample includes the production of the $\psip$ resonance,
       the ISR production of the $J/\psi$, and the continuum processes
       incorporated in {\sc kkmc}~\cite{ref:kkmc}. All particle decays are modeled
       with {\sc besevtgen}~\cite{ref:evtgen,ref:bes3_generators} using branching fractions either taken from
       the particle data group (PDG)~\cite{pdg}, when available, or otherwise
       estimated with {\sc lundcharm}~\cite{ref:lundcharm1,ref:lundcharm2}.
       However, $\chi_{c(0,1,2)} \to 3\pi^0$ decays produced by the {\sc lundcharm} model
       are omitted from the analysis, because such decays are expected to be strongly suppressed by 
       either $P$-parity or $G$-parity conservation.
       Final state radiation from charged
       particles is incorporated using the {\sc photos} package~\cite{ref:photos}.

       In total, $2748$ million inclusive MC events are used for background evaluation and
       $13$ million signal MC events generated via the phase space model (PHSP MC) are used for
       the efficiency evaluation and for the reweighing procedure during the amplitude analysis.

\section{Event selection \label{sec:ev_sel}}

       The final state for the signal events consists of five photons: one
       from radiative decay of $\psip$ and two pairs from $\pi^0$ decays.
       Therefore, candidate events are required to contain no charged tracks and at least
       five good photon candidates. Each photon is required to have an energy deposition
       of at least $50$~MeV in the EMC, regardless if it is reconstructed in the barrel
       or end cap region.
       For additional suppression of electronic noise in the EMC and showers
       which are unrelated to the event, the timing for each shower is required
       lie within $\pm500$~ns of the most energetic shower.

       After the above selection, we apply a 6C kinematic fit to all possible $\gamma\pipi$
       permutations to increase the accuracy of the kinematic parameters of the event
       and, as a consequence, allow for better suppression of the combinatorial background.
       The procedure is performed in two stages. First, a 4C kinematic fit is applied to all possible
       groups of five photons. The four constraints enforce four-momentum conservation between the final state of $5\gamma$
       and the initial state of $\psip$. The set of $5\gamma$ with the smallest $\chi^2_{\rm{4C}}$
       is then subjected to a 6C kinematic fit.
       In addition to the four-momentum constraints, it imposes two constraints
       for each photon pair requiring an invariant mass equal to the mass of $\pi^0$.
       The 6C fit is applied to all possible permutations of the five photons.
       If multiple permutations survive the selection,
       only the one with the lowest $\chi^2_{\rm{6C}}$ is retained.
       The value of $\chi^2_{\rm{6C}}$ is required to be below 35.
       Additionally, $M_{\text{inv}}(\pipi)<3$~\GeVcc\ is applied to to focus on the mass region relevant for the amplitude analysis.

       Additional selections are applied to suppress the most notable background contributions.
       In the context of this analysis, $\psip \to \omega \pi^0 \to \gamma \pipi$
       and $\psip \to b_1^0 \pi^0 \to \gamma \pipi$, are treated as
       background processes. Despite having the same final state, they are not
       radiative $\psip$ decays and therefore
       do not probe the dynamics of the $\pipi$ system.
       To suppress background from $\psip$ decays to $\omega \pi^0$,
       the invariant mass of each $\gamma \pi^0$ combination
       is required to satisfy:
       $|M_{\text{inv}}(\gamma \pi^0_{(1,2)}) - M_\omega| > 50$~\MeVcc,
       where the $\omega$ mass is taken from PDG~\cite{pdg}.
       Due to the large
       width of the $b_1^0$ resonance, efficient suppression of the $\psip\to b_1^0 \pi^0 $ is not feasible. The impact of this background
       is studied as a source of systematic uncertainty.

       \begin{figure*}[htbp]
       \includegraphics[width=0.98\textwidth]{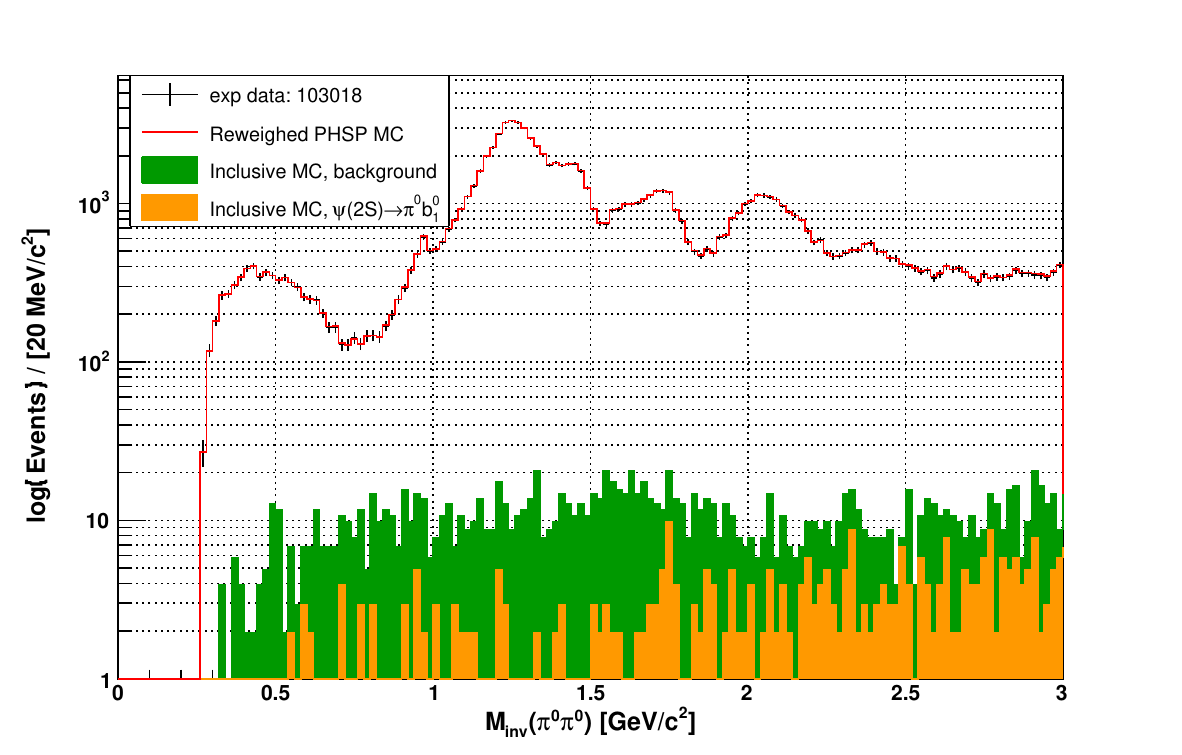}
       \caption{
       Spectrum of the $M_{\text{inv}}(\pipi)$ after all selection procedures have been applied.
       Black crosses represent experimental data with statistical errors,
       while the red line shows selected PHSP MC
       events which are reweighed using solutions from the amplitude analysis.
       The green area shows the total background estimated using the
       inclusive MC sample and the orange area specifically denotes the
       $\psip \to b_1^0 \pi^0$ events in it.
       \label{fig:inv_mass_spectrum}
       }
       \end{figure*}

       Signal events in which an incorrect association of photons has occurred
       represent another important background source
       due to the significant distortion of its kinematics.
       We call these events misreconstructed. They mainly
       arise from incorrect assignment of the radiative photon to one of the $\pi^0$ photons. To suppress this type of background,
       the invariant mass of the radiative photon paired with any one
       $\pi^0$ daughter photon must exceed $175$~\MeVcc.

       A total of $103,018$ events from the initial data sample have passed all selection criteria.
       Figure~\ref{fig:inv_mass_spectrum} shows the $\pipi$ invariant mass spectrum.
       A comprehensive background study is performed to evaluate the purity of the selected sample.
       Background processes with final state different from $5\gamma$ and
       the $\psip \to b_1^0 \pi^0 \to \gamma \pipi$ background
       are evaluated using the inclusive MC sample described above.
       The resulting inclusive MC background contribution is estimated to be $(0.92\pm0.03)\%$.

       An additional data-driven estimate of
       the $\gamma\pi^0\pi^0\pi^0$ background is performed.
       The selection procedure is similar to the nominal selection, with the primary difference being the use of a 7C kinematic fit instead of the 6C fit to account for the additional $\pi^0$ in the final state.
       After correcting for differences in selection efficiencies, the contribution of this background is conservatively estimated to be $0.5\%$.
       Possible continuum background contributions are investigated using
       the off-resonance data sample at $\sqrt{s}=3650$~MeV,
       yielding an estimated contribution of around $0.2\%$.
       The fraction of misreconstructed events is estimated using the PHSP MC sample. An event is classified as misreconstructed if its reconstructed kinematics differ significantly from the MC truth information. Each misreconstructed PHSP MC event is taken into account
       with its weight defined as a ratio of the $M_{\text{inv}}(\pipi)$ spectra observed in data and PHSP MC.
       The level of misreconstructed background is estimated to be $(0.14\pm0.01)\%$.

       As an independent validation of the background obtained using the inclusive MC data, an
       additional sideband study was performed using a modified selection procedure.
       This selection includes the charged-track veto, the same photon requirements as the nominal selection,
       the 4C kinematic fit and the suppression of
       $\psip\to\omega \pi^0$ and misreconstructed events.
       The $M_{\text{inv}}(\pipi)$ spectra were studied in two-dimensional sidebands
       in the $M_{\text{inv}}(\gamma\gamma) \times M_{\text{inv}}(\gamma\gamma)$ plane.
       Good agreement is observed between experimental data and inclusive MC simulation, indicating that the inclusive MC sample provides an adequate description of the residual background. The remaining discrepancies are generally small and do not significantly exceed twice the estimated inclusive MC background level. These differences are conservatively treated as an            additional source of systematic uncertainty.

\section{Amplitude analysis \label{sec:pwa}}

       Amplitude analyses, also called partial-wave analyses, or PWAs for short,
       have long been an essential tool in hadron spectroscopy (see, e.g., Ref.~\cite{Anisovich:2008zz}).
       Their usage is inevitable when disentangling overlapping resonances and their interference effects.
       Usually such analyses are carried out by factorizing the decay amplitude into a dynamic part
       and an angular part. The latter one can be expressed in several equivalent formalisms;
       however, the former is usually much more important because it describes the dynamics of particle
       interactions.

       There are two main approaches to parametrize the dynamic part of the amplitude.
       The first approach is to provide parametrization using some underlying model; for example,
       it could be a coherent sum of resonances parameterized by Breit-Wigner functions, or
       the $K$-matrix parametrization. Analyses of this type, which are usually called model-dependent,
       have a clear advantage that the model parameters can be directly linked to properties
       of the scattering amplitude in the complex $s$-plane; here $s$ is the Mandelstam variable.

       The second approach is to measure the dynamical amplitude as a function of invariant mass
       by performing the fit procedure in each invariant mass bin
       while making minimal model assumptions. Under such model-independent approach,
       the dynamic part is simply taken as a constant for each partial wave in each bin.
       This approach remains applicable even when conventional resonance parametrizations become inadequate, particularly in cases involving broad and overlapping resonances. The results of
       such analyses are useful for the subsequent development of more complex models for which
       the results could be a base. Another important benefit is that results of multiple
       model-independent analyses could be included in a global fit at the same time.

       Both the benefits and the drawbacks of both approaches are discussed in literature
       extensively, especially in Ref.~\cite{PWA_Jpsi_pipi} and Ref.~\cite{PWA_Jpsi_KsKs}.
       Due to the complexity of the light hadron spectrum, direct model-dependent analysis
       of the selected data sample would provide results of very limited value and applicability
       outside of strong model assumptions.
       Therefore, in this study we focus on the model-independent (MI) approach instead.

       \subsection{Model-independent amplitude analysis formalism \label{sec:pwa:mi_formalism}}
       The general formalism of the MI analysis is the same as the one
       described in Ref.~\cite{PWA_Jpsi_pipi} in details.
       Amplitudes for radiative $\psip$ decays to $\pipi$ are obtained in the radiative multipole basis.

       The quantum-mechanical amplitude of $\psip$ decaying to $\pipi$ is written as:
       \begin{equation}
       U^{M,\lambda_\gamma}(\vec{x},s) = \braket{\gamma \pipi|H|\psip},
       \end{equation}
       where $s$ is the squared invariant mass of the $\pipi$ system,
       $\vec{x} = (\theta_\gamma, \phi_\gamma, \theta_\pi, \phi_\pi)$
       are the remaining parameters to describe the point in phase space,
       $H$ is a Hamiltonian which describes the transition,
       $M$ is the $\psip$ polarization and $\lambda_\gamma$ is the radiative photon's
       helicity. For our process, possible values for both $M$ and $\lambda_\gamma$ are $\pm1$.
       This amplitude is factorized into two parts, one describing a radiative transition
       into the intermediate state $X$ and the other describing the strong decay of $X$ into the $\pipi$ pair:
       \begin{multline}
       \label{eq:amplitude_qcd_em_par}
       U^{M,\lambda_\gamma}(\vec{x},s) = \sum_{j, J_\gamma, X}
       \braket{\pipi|H_{\text{QCD}}|X_{j,J_\gamma}} \times \\
       \braket{\gamma X_{j,J_\gamma}|H_{\text{EM}}|\psip},
       \end{multline}
       where $j$ is the angular momentum of the intermediate state $X$ and
       $J_\gamma$ is the radiative photon angular momentum,
       which also indexes radiative multipole transitions.
       Summation over $X$ includes possible pseudoscalar-pseudoscalar states which could rescatter
       into $\pipi$. These rescattering effects become important for the $M_{\text{inv}}(\pipi)$ values
       above $KK$ threshold.

       From the Eq.~\ref{eq:amplitude_qcd_em_par}, 
       the decay amplitude may be further factorized into the following:
       \begin{equation}
       U^{M,\lambda_\gamma}(\vec{x},s) = \sum_{j,J_\gamma}
       V_{j,J_\gamma}(s) A_{j,J_\gamma}^{M,\lambda_\gamma}(\vec{x}).
       \end{equation}
       Here $A_{j,J_\gamma}^{M,\lambda_\gamma}(\vec{x})$ describes the angular dependencies and
       is fully determined by event kinematics, while $V_{j,J_\gamma}(s)$ is called a coupling
       to the state with $(j,J_\gamma)$ quantum numbers and describes the $\pipi$ dynamics.

       Amplitudes with non-zero total angular momentum will have three different components,
       while the $0^{++}$ amplitude will have only one component.
       The baseline fit includes the $0^{++}$ E1 amplitude 
       and the $2^{++}$ E1, M2 and E3 amplitudes.
       Theoretically, any amplitude with even total angular momentum and positive parity
       and charge conjugation is accessible for the studied decay.
       However, the study of the baseline fit quality shows that there are
       no $M_{\text{inv}}(\pipi)$ bins for which discrepancy in angular distributions
       between experimental data and model with couplings obtained from the fit exceeds $4\sigma$.
       Additionally, there are only 4 out of 137 bins for which discrepancy exceeds $3\sigma$.
       Thus, no statistically significant improvement in fit quality is observed with
       the addition of $4^{++}$ and higher amplitudes.

       In the MI analysis, the data sample is binned as a function of the $\pipi$ invariant mass.
       Within each narrow mass bin, the partial-wave couplings are assumed to be
       constant, which allows to avoid strong model dependent assumptions about the
       dynamical part of the decay amplitude:
       \begin{equation}
       U^{M,\lambda_\gamma}(\vec{x},s) = \sum_{j,J_\gamma}
       V_{j,J_\gamma} A_{j,J_\gamma}^{M,\lambda_\gamma}(\vec{x}).
       \end{equation}
       The amplitudes are extracted independently from the fit procedure performed in each bin
       of the $\pipi$ invariant mass. The fit procedure itself is the unbinned extended
       maximum likelihood fit with couplings taken as free parameters and obtained as a result.
       The final set of couplings allows to calculate partial-wave intensities in each bin,
       which describe the dynamics of the $\pipi$ interaction.

       The overall intensity function for the PHSP point $\vec{x}$ is written as:
       \begin{equation}
       I(\vec{x}) = \sum_{M,\lambda_\gamma}
       \left|
       \sum_{j, J_\gamma} V_{j,J_\gamma} A_{j,J_\gamma}^{M,\lambda_\gamma}(\vec{x})
       \right|^2,
       \end{equation}
       where the incoherent sum includes the non-measured observables of the reaction.
       In this case they are the polarization of the $\psip$, $M = \pm1$, and the helicity
       of the radiative photon, $\lambda_\gamma = \pm1$. The free parameters of the fit
       are assumed to be independent from these observables.

       In the results that follow we show intensities for the specific partial waves in each bin,
       which are reported as a number of events corrected for the detector efficiency and acceptance.
       The intensity in the bin $k$ (bound by $s_k$ and $s_{k+1}$) for the partial wave
       with quantum numbers $(j,J_\gamma)$ is given by:
       \begin{equation}
       I^{k}_{j,J_\gamma} = \int_{s_k}^{s_{k+1}}\sum_{M,\lambda_\gamma}
       \left|
       V_{j,J_\gamma} A_{j,J_\gamma}^{M,\lambda_\gamma}(\vec{x})
       \right|^2 d\vec{x}.
       \end{equation}
       In the actual fit procedure, the PHSP volume is absorbed into the free
       parameters. As a result, the actual fit parameters $V'_{j,J_\gamma}$ are a product
       of the coupling $V_{j,J_\gamma}$ and the square root of the PHSP volume in the bin $k$.

       \subsection{Ambiguities \label{sec:pwa:ambiguities}}
       The MI analysis has an additional set of complications which arise due to
       the presence of both trivial and non-trivial ambiguities.
       Trivial ambiguities arise due to the freedom to rotate the overall amplitude by $\pi$
       or to complex conjugate the overall amplitude without it resulting in the change
       of the partial-waves intensities in the chosen bin. These ambiguities are
       removed by applying additional phase conventions.
       The source of non-trivial ambiguities is described in detail in Ref.~\cite{PWA_Jpsi_pipi}.
       In short, amplitudes with the same quantum numbers may have
       different phases above the $KK$-threshold, where rescattering processes become impactful.
       For the radiative decays of $\psip$ into the pair of pseudoscalars, only two
       solutions are present if $0^{++}$ and $2^{++}$ amplitudes are taken into account.

       \subsection{Background subtraction \label{sec:pwa:bkg_sub}}

       The remaining background contamination of the selected data sample
       is taken into account by the subtraction procedure via the MC sample.
       As shown in Ref.~\cite{PWA_Jpsi_pipi}, the likelihood function for the extended maximum
       likelihood fit with background subtraction may be written as following:
       \begin{equation}
       L(\vec{\xi}) = \frac{e^{-\mu}\mu^{N_{\text{data}}}}{N_{\text{data}}!}
       \prod_{i=1}^{N_{\text{data}}} f(\vec{x}_i|\vec{\xi})
       \prod_{j=1}^{N_{\text{bkg}}^{\text{MC}}} f(\vec{x}_j|\vec{\xi})^{-w_j}.
       \end{equation}
       Here $f(\vec{x}|\vec{\xi})$ is the probability density function
       to observe an event with a set of kinematics $\vec{x}$
       and fit parameters $\vec{\xi} = \{ V_{j,J_\gamma}^{k}\}$, $N_\text{data}$ is the size of the
       selected data sample, $N_{\text{bkg}}^{\text{MC}}$ is the
       size of the MC sample used to approximate
       remaining background, $w_j$ is the weight used for scaling purposes.
       We use this approach to subtract non-$5\gamma$ and $b_1^0 \pi^0$
       backgrounds estimated via the inclusive MC sample.

       However, if background contamination is small, then the background
       can be neglected altogether. In this analysis, $\gamma3\pi^0$ final state,
       misreconstructed and nonresonant background
       levels are much lower than the inclusive MC one and are sufficiently small to be neglected.

       \subsection{Statistical uncertainties \label{sec:pwa:stat_uncertainties}}

       The main result of this analysis is a set of partial-wave intensities
       obtained in each bin. However, the strongly non-parabolic likelihood behavior exhibited by the log-likelihood
       function near the minima does not allow to define statistical
       uncertainties applicable for subsequent phenomenological analyses of results.
       In this situation, quoting statistical uncertainties as single symmetric values becomes insufficient.

       Because of the aforementioned reasons,
       statistical information is provided in a form of negative log-likelihood scans
       (NLL scans for short). In this context, NLL scans are defined as functions
       of $-\ln{(L)}$ versus the corresponding partial-wave intensity
       $I^{k}_{j,J_\gamma}=\{I^{k}_{01},I^{k}_{21},I^{k}_{22},I^{k}_{23}\}$ in the mass bin $k$.
       Each partial-wave intensity
       is linked to the respective coupling via the following relation:
       \begin{equation}
       I^{k}_{j,J_\gamma} = C^k \times \left|V^{k}_{j,J_\gamma}\right|^2,
       \end{equation}
       where $C^k$ is a constant which depends only on the bin number and partial-wave quantum
       numbers, and is independent from the coupling. Therefore, a fixed coupling results in 
       a fixed partial-wave intensity, which in turn allows for obtaining likelihood scans 
       as functions of partial-wave intensities rather than couplings.

       As discussed in Ref.~\cite{ref:likelihood_to_significance}, variation of $-2\ln(L)$ by $N^2$
       from the minimum allows to obtain the $N$-standard deviation confidence interval
       for partial-wave intensities.
       This allows us to obtain $1\sigma$, $2\sigma$ and $3\sigma$ curves, which are
       plotted along with nominal intensity values to illustrate the obtained results.

\begin{figure*}[htbp]
       \begin{center}
       \includegraphics[width=\textwidth]{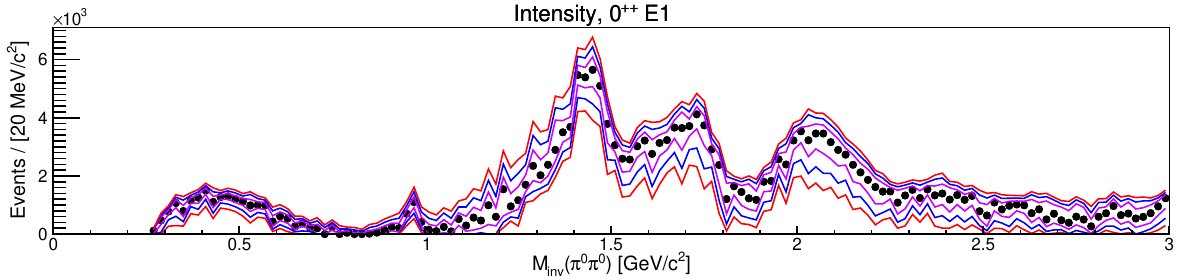}
       \includegraphics[width=\textwidth]{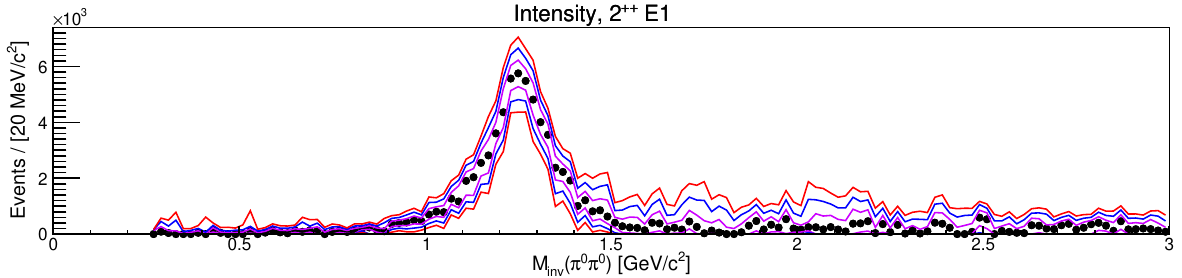}
       \includegraphics[width=\textwidth]{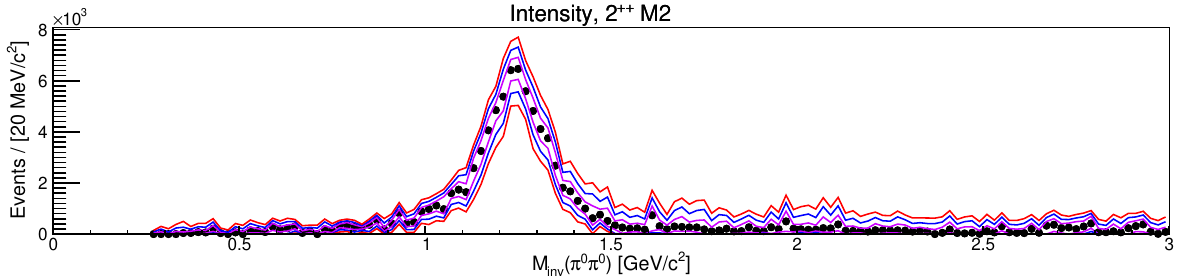}
       \includegraphics[width=\textwidth]{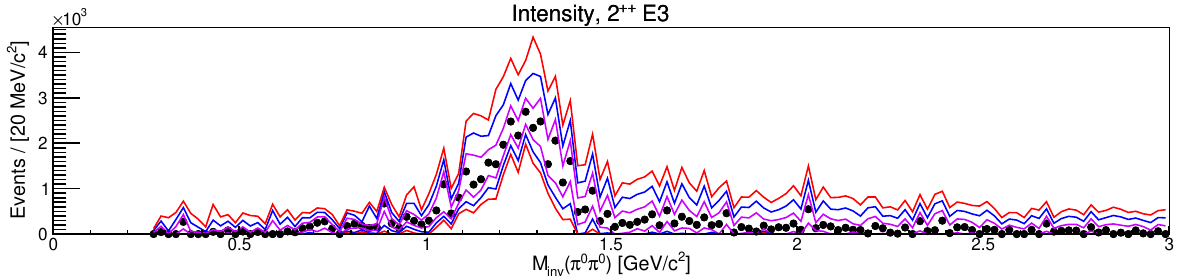}
       \caption{
               Partial-wave intensities. Black dots are intensity values obtained from scans
               modified by the convolution procedure to include Type~2 systematic uncertainties.
               The colored graphs correspond to changes
               of $-2\ln(L)$ by $1$, $4$ and $9$, which are generally interpreted as
               the uncertainties of $1\sigma$, $2\sigma$ and $3\sigma$, respectively.
               Statistical uncertainties along with Type~2 systematic uncertainties are shown.
       \label{fig:int_graphs}
       }
       \end{center}
       \end{figure*}

       Because of the convolution procedure applied during the systematic uncertainties estimation,
       the shape of the NLL scans is changed from the initial ones.
       It results in changes in $N\sigma$ curves,
       as well as breaking of the ambiguity between the two solutions. Due to that, the final results are shown
       in Fig.~\ref{fig:int_graphs} after systematic uncertainties are discussed.

       \subsection{Systematic uncertainties \label{sec:pwa:syst_uncertainties}}

       The main source of systematic uncertainties for this analysis are uncertainties
       related to the overall normalization of the results. Systematic uncertainties are divided into two categories,
       which we call Type~1 and Type~2.

       The impact from Type~1 sources is estimated via
       a bin-independent factor without the need of propagation
       through the PWA procedure. These sources are the photon detection efficiency,
       the total number of $\psip$, backgrounds with the $\gamma 3\pi^0$ final state and
       nonresonant backgrounds. Photon detection efficiency was studied in Ref.~\cite{rec_neutral_tracks},
       uncertainty from the differences in neutral particle reconstruction between data and MC
       does not exceed $1\%$ per photon. The number of $\psip$ in the studied data was
       determined in Ref.~\cite{Ablikim_2024} to be $N_{\psip} = (2712.4 \pm 14.3)\times 10^6$,
       with an uncertainty of $0.5\%$. We attribute conservative estimation of
       systematic uncertainty equal to $100\%$ of the background contamination
       for the backgrounds with $\gamma3\pi^0$, with uncertainty from
       this source of $0.5\%$. The same is performed for the nonresonant background contributions,
       with resulting uncertainty of $0.2\%$. The uncertainty of the branching
       fraction of $\pi^0 \to \gamma\gamma$ decays is $0.03\%$ according to the PDG~\cite{pdg}
       and can be neglected. The sources of Type~1 systematic uncertainties are
       summarized in Table~\ref{tab_syst_summary}.
       As one would notice, the largest impact on the overall uncertainty is due to
       the photon detection efficiency. Overall systematic uncertainty from
       Type~1 sources is $5.1\%$.

       \begin{table}[t]
       \centering
       \caption{
               Type~1 systematic uncertainties for the overall normalization
               of the partial-wave intensities.
       }
       \label{tab_syst_summary}
       \begin{tabular}{cc}
               \hline \hline
               Systematic uncertainty source & $\psip \to \gamma \pipi,~\%$ \\
               \hline
               Photon detection efficiency & $5.0$ \\

               Number of $\psip$ & $0.5$ \\

               Backgrounds with $\gamma 3\pi^0$ final state & $0.5$ \\

               Nonresonant backgrounds & $0.2$ \\
               \hline
               \hline
               Total from Type~1 sources & 5.1 \\
               \hline
               \hline
       \end{tabular}
       \end{table}

       The impact of Type~2 sources is bin-dependent and therefore can not be estimated
       with a single number. Instead, it is measured via propagation through the PWA procedure
       and then modifying NLL scans to include these uncertainties.
       Type~2 sources are uncertainties due to limits of the 6C kinematic fit,
       non-$5\gamma$ backgrounds from inclusive MC
       and events with the $b_1^0 \pi^0$ intermediate state. For each of the sources
       an alternative selection procedures are applied and the whole PWA procedure is repeated.
       The limit on the $\chi^2_{\rm{6C}}$ is changed to $50$ from the nominal value of $35$,
       subtracted non-$5\gamma$ backgrounds are varied by $100\%$ and $\psip \to b_1^0 \pi^0$
       events are suppressed using the following cuts:
       \begin{equation}
       \begin{split}
               |M_{\text{inv}}(\gamma,\pi^0_1) - M_{b_1^0}| &>70~\text{\MeVcc},\\
               |M_{\text{inv}}(\gamma,\pi^0_2) - M_{b_1^0}| &>70~\text{\MeVcc}.
       \end{split}
       \end{equation}
       For each of this three sources, the difference in the total intensity in each bin is taken as
       a parameter $\sigma_{i}^{k}$, where $i$ is the source number and $k$ is the bin number.
       Total parameter $\sigma^{k}_{\text{total}} = \sqrt{\sum_{i=1}^{3} {\sigma_{i}^{k}}^2}$
       is then used to modify NLL scans in the bin number $k$ to take Type~2 systematic
       uncertainties into account. This procedure is performed for all bins and is described
       below.

       The Type~2 systematic uncertainties are taken into account in each bin via
       the convolution of NLL scans with a Gaussian function with
       the variance $\sigma \equiv \sigma^k$ defined earlier.

       Nominal NLL scans are graphs of $-\ln(L)$ versus $I_{j,J_\gamma}$.
       From them, using linear interpolation, we obtain functions
       $-\ln(L(I_{j,J_\gamma})) \equiv -\ln(L_{j,J_\gamma}(x))$.
       The modified likelihood function is calculated using the following relation:
       \begin{widetext}
       \begin{equation}
       L_{j,J_\gamma}^{\text{mod}}(x) = \int_{x-3\sigma}^{x+3\sigma}{
       \frac{1}{\sqrt{2 \pi \sigma^2}} \exp{\left(-\frac{(x'-x)^2}{2\sigma^2}\right)} L_{j,J_\gamma}(x') dx'
       }
       /
       \int_{0}^{x+3\sigma}{
       \frac{1}{\sqrt{2 \pi \sigma^2}} \exp{\left(-\frac{(x'-x)^2}{2\sigma^2}\right)} dx'
       }.
       \end{equation}
       \end{widetext}

       The second integral in this formula provides the correction needed due to the fact that
       the partial-wave intensity is a non-negative value. Therefore, any smearing caused by convolution
       with a Gaussian function must not result in negative values of intensities.

       Finally, modified scans are obtained by calculating values of $-\ln(L_{j,J_\gamma}^{\text{mod}}(x))$
       for the same values of $x = I_{j,J_\gamma}$ as original scans.
       This procedure results in broadening of valleys
       near minimum, which in term increases values of statistical uncertainties
       inferred from scans.

       However, the convolution results in the second, more profound impact on partial-wave intensities. Because of how the procedure is constructed, it is sensitive not only
       to the values of log-likelihood function in its minima, but also to the shape of it.
       Because the topology of likelihood function is different near each of the minima,
       it ends up lifting the degeneracy between the two ambiguous solutions. Only one of them ends up as a new global
       minimum and is used as the final result for the respective partial-wave intensity.

       \subsection{Results \label{sec:pwa:res}}

       The intensities for each amplitude are depicted in Fig.~\ref{fig:int_graphs}
       along with curves which represent $1\sigma$, $2\sigma$ and $3\sigma$
       uncertainty intervals in each bin. Due to the convolution used to take Type~2
       systematic uncertainty into account, the ambiguity between solutions is broken
       and only one nominal intensity per respective amplitude remains in each bin.

       A similar procedure is performed for obtained phase differences between partial waves.
       After the convolution procedure is applied to take into account
       Type~2 systematic uncertainties, phase differences for all waves
       have enough statistically significant information
       only in the narrow region $1.0-1.5$~\GeVcc.
       The phase difference is considered to be statistically significant if
       its estimated $3\sigma$ confidence interval is less than $2\pi$.
       This makes obtained phase differences of little use in the context
       of subsequent studies.

       Decay patterns observed in the $0^{++}$ intensity are very similar
       to the ones observed in the previous model-independent $J/\psi$ study~\cite{PWA_Jpsi_pipi}.
       Additionally, in the $2^{++}$ waves most points
       above $1.5$~\GeVcc\ are within $1\sigma$ from zero,
       which means that we do not observe any evidence
       for the existence of the tensor glueball in the studied mass range
       from the partial-wave intensities alone.
       Nevertheless, obtained results
       of the model-independent amplitude analysis allow for their further study
       within more constrained model-dependent frameworks
       and provide valuable dataset relevant for future analyses.

       \section{Conclusions \label{sec:conclusions}}
       A mass-independent amplitude analysis of the $\pipi$ system produced
       in the radiative $\psip$ decays is presented. Piecewise descriptions of the scalar and tensor partial-wave intensities are obtained as
       a result of processing the largest $\psip$ decay dataset available up to date.
       The strategy used in the analysis allows to minimize systematic bias from
       $\pipi$ dynamics assumptions, while simultaneously allowing subsequent
       development of models which parametrize data obtained in this analysis,
       either by itself or in conjunction with other datasets.

               While the contribution of the scalar partial wave exhibits complex
       pattern of multiple overlapping resonances, the tensor contributions
       provide no statistically significant evidence for resonance-like
       structures except for the clearly visible $f_2(1270)$. Thus, no
       clear evidence for the tensor glueball in the studied mass range could be obtained
       from the partial-wave intensities alone.

       The complicated non-parabolical behavior of likelihood scans
       near minima led to an alternative method of statistical uncertainties'
       estimation. The method was further extended to
       take into account the complicated mass-dependent systematic
       uncertainties, while at the same time resolving the ambiguity between
       multiple solutions commonly present in the mass independent analyses.

       In order to allow for model development, the results of the analysis
       are provided in two different ways. Firstly, they are shown here as a function
       of $M_{\text{inv}}(\pipi)$. Secondly, parametrization of likelihood scans for each
       partial amplitude in each bin are provided in supplemental materials.

\begin{acknowledgments}
The BESIII Collaboration thanks the staff of BEPCII (https://cstr.cn/31109.02.BEPC) and the IHEP computing center for their strong support. This work is supported in part by National Key R\&D Program of China under Contracts Nos. 2025YFA1613900, 2023YFA1606000, 2023YFA1606704, 2025YFA1613900; National Natural Science Foundation of China (NSFC) under Contracts Nos. 11635010, 11935015, 11935016, 11935018, 12025502, 12035009, 12035013, 12061131003, 12192260, 12192261, 12192262, 12192263, 12192264, 12192265, 12221005, 12225509, 12235017, 12342502, 12361141819, 12535005; the Chinese Academy of Sciences (CAS) Large-Scale Scientific Facility Program; the Strategic Priority Research Program of Chinese Academy of Sciences under Contract No. XDA0480600; CAS under Contract No. YSBR-101; 100 Talents Program of CAS; The Institute of Nuclear and Particle Physics (INPAC) and Shanghai Key Laboratory for Particle Physics and Cosmology; Agencia Nacional de Investigación y Desarrollo de Chile (ANID),
Chile under Contract No. ANID CCTVal CIA250027; ERC under Contract No. 758462; German Research Foundation DFG under Contract No. FOR5327; Istituto Nazionale di Fisica Nucleare, Italy; Knut and Alice Wallenberg Foundation under Contracts Nos. 2021.0174, 2021.0299, 2023.0315; Ministry of Development of Turkey under Contract No. DPT2006K-120470; National Research Foundation of Korea under Contract No. NRF-2022R1A2C1092335; National Science and Technology fund of Mongolia; Polish National Science Centre under Contract No. 2024/53/B/ST2/00975; STFC (United Kingdom); Swedish Research Council under Contract No. 2019.04595; U. S. Department of Energy under Contract No. DE-FG02-05ER41374

\end{acknowledgments}

\appendix
\section{Supplemental materials \label{appendix_how_to_use_our_results}}
       In this appendix we describe how these results could be used in other external analyses.

       Assume that a model-dependent analysis is performed and results of this analysis are
       being included in the global fit. Necessary, albeit expected assumption here is that model used
       in the analysis predicts partial-wave intensities as a function of energy $I_{\alpha}(s)$, where
       $\alpha \equiv (j,J_\gamma) \in \{(0,1),(2,1),(2,2),(2,3)\}$ is an index for partial wave and
       $s$ is the energy squared in CMS. Then part of $\ln(L)$ which is based on our data in global fit
       could be calculated using following formula:
       \begin{equation}
       \ln(L_{\psip \to \gamma \pipi}) = \sum_{k=1}^{150} \sum_{\alpha}
       L^{k}_{\alpha}( I_{\alpha}(s_k) ),
       \end{equation}
       where $k$ is the number of bin in the $M_{\text{inv}}(\pipi)$ spectrum,
       $s_k = 0.01 + (k-1) \times 0.02$~$\text{\GeVcc}$ is
       the center of the respective bin and $L^{k}_{\alpha}$ is the scan for respective partial-wave intensity
       in the bin number $k$.

       In the supplemental materials we provide a set of text files
       which contain NLL intensity scans~\cite{ref:supplemental_materials}.

\bibliography{paper_ref.bib}

\begin{thebibliography}{41}
\bibitem [{ref()}]{ref:supplemental_materials}%
  \BibitemOpen
  \href@noop {} {}\bibinfo {note} {See [URL to be inserted by the publisher]
  for the full likelihood scans for all four partial waves.}\BibitemShut
  {Stop}%
\end{thebibliography}

\end{document}